\documentclass[a4paper,11pt]{article}
\usepackage{pos}

\title{Kaon physics in the SMEFT}

\author[a]{Jason Aebischer}

\affiliation[a]{PSI Center for Neutron and Muon Sciences,\\
  5232 Villigen PSI, Switzerland}

\emailAdd{jason.aebischer@psi.ch}

\abstract{Kaon physics observables are highly sensitive to New Physics (NP) effects and form in combination with the Standard Model Effective Field Theory (SMEFT) a powerful tool to study physics that goes beyond the Standard Model paradigm. We review recent SMEFT analyses in the Kaon sector and point out novel directions that might be investigated in the future.
}

\FullConference{
XIII International Conference on Kaon Physics -- Kaon 2025 Mainz
}

\begin{document}
\maketitle

\section{Introduction}

Kaon physics provides one of the most sensitive means to study effects that go beyond the Standard Model (SM) of particle physics. The discovery of CP violation in the Kaon sector via the Cronin-Fitch experiment \cite{Christenson:1964fg} or the prediction of the charm quark from $K^0-\bar K^0$ mixing through the GIM mechanism \cite{Glashow:1970gm} are only two of the milestones achieved from Kaon physics that lead to a deeper understanding of the SM as we know it today. An important class of processes in the search for BSM physics are rare Kaon decays such as $K\to \pi \nu \bar{\nu}$ and $K\to \pi \ell \bar{\ell}$, which are sensitive to very high scales and therefore allow to test New Phyics (NP) effects way beyond the reach of current colliders. The interest in these decays was corroborated further by the latest measurement of the golden mode by the NA62 collaboration \cite{NA62:2024pjp}
\begin{equation}
\text{BR}(K^+\to \pi^+ \nu \bar{\nu})_{\text{exp}} = (13.0^{+3.3}_{-3.0} )\times 10^{-11}\,,
\end{equation}
and an update of this value is expected at the end of LS3 of the LHC. An upper bound on the rare neutral decay mode is given by the KOTO collaboration \cite{KOTO:2024zbl}
\begin{equation}
\text{BR}(K_L\to \pi^0 \nu \bar{\nu})_{\text{exp}} < 2.2 \times 10^{-9} \quad (90\% \text{ C.L.})\,,
\end{equation}
and a measurement of this decay is expected in the 2030s at the KOTO-II experiment \cite{KOTO:2025gvq}. Furthermore, the Belle II collaboration reported evidence for the decay $B^+\to K^+ \nu \bar{\nu}$ with a significance of 3.5$\sigma$ \cite{Belle-II:2023esi}. All these measurements allow to test a plethora of NP scenarios. However, instead of investigating multiple NP models individually, it is more efficient to adopt an Effective Field Theory (EFT) approach. One of the most widely used EFT frameworks is the Standard Model Effective Field Theory (SMEFT), which provides a systematic and model-independent framework to describe heavy NP, assuming invariance under $SU(3)_C\times SU(2)_L\times U(1)_Y$ as well as Poincar\'e transformations.\footnote{For recent reviews on the SMEFT see \cite{Brivio:2017vri,Isidori:2023pyp,Aebischer:2025qhh}.} In the following we provide an overview over recent SMEFT analyses conducted in the Kaon sector and point out possible avenues for future investigation.

\section{Rare Kaon decays}
Rare Kaon decays such as $K^+\to \pi^+ \nu \bar{\nu}$, $K_L\to \pi^0 \nu \bar{\nu}$ and also $K_S\to \mu^+\mu^-$ were studied in the SMEFT for instance in \cite{Aebischer:2020mkv} in the context of $Z'$ models. It was shown that for a suppression of $\Delta M_K$ either left-handed or right-handed flavor violating $Z'$ couplings were needed, which imply strong correlations among rare modes. A similar scenario was studied in \cite{Aebischer:2023mbz} under the assumption that there is no NP in $\epsilon_K$. In this case only left-handed imaginary $Z'$ couplings were found to satisfy all constrains, while predicting a strongly correlated pattern for the rare decays. 

Collider constraints involving rare Kaon decays were studied in \cite{Roy:2024avj}, finding that the constraints on SMEFT operators from flavor physics are still stronger than the current reach of colliders. The possibility of NP effects in the third generation was studied in \cite{Allwicher:2024ncl}, by imposing a $U(2)^5$ flavor symmetry on the first two generations. In this scenario, bounds on semileptonic SMEFT Wilson coefficients as well as flavor violating parameters were studied, by taking into account constraints from $K^+\to \pi^+ \nu \bar{\nu}$ and $B^+\to K^+ \nu \bar{\nu}$. In \cite{Marzocca:2024hua} the latter observable together with electroweak (EW) and collider constrains was used to constrain SMEFT operators as well as different scalar leptoquark scenarios. Furthermore, constraints from rare Kaon decays and their impact on various semileptonic SMEFT operators were studied in detail in \cite{Kumar:2021yod}. At the dimension-seven level lepton number violating operators and their impact on differential distributions of rare decays were studied in \cite{Deppisch:2020oyx}, probing the Majorana nature of neutrinos. A general analysis contrasting Majorana to Dirac type neutrinos in $K^+\to \pi^+ \nu \bar{\nu}$ distributions can be found in \cite{Gorbahn:2023juq}. The possibility to distinguish NP effects in $q^2$-distributions of rare decays were worked out for the vector case in \cite{Li:2019fhz}, for scalar NP in \cite{Deppisch:2020oyx} and in more generality in \cite{Buras:2024ewl}. Finally, possible SMEFT effects occurring in $q^2$-distributions of rare Kaon decays were studied in \cite{Aebischer:2022vky}.

\section{$\epsilon'/\epsilon$}
One of the prime examples to study direct CP violation is the observable $\epsilon'/\epsilon$, for which interference of isospin amplitudes yields CP violation in decay. An analysis on the SM prediction based on recent lattice results \cite{RBC:2020kdj} for the hadronic matrix elements can be found in \cite{Aebischer:2020jto}. Concerning SMEFT analyses a master formula for this observable was derived in \cite{Aebischer:2018csl}, where all relevant SMEFT operators as well as the complete SMEFT running \cite{Alonso:2013hga,Jenkins:2013wua,Jenkins:2013zja} above the EW scale was taken into account, using the Python package \texttt{wilson} \cite{Aebischer:2018bkb}. The findings are based on the results in \cite{Aebischer:2018quc}, where the running below the EW scale was taken into account. For the SM hadronic matrix elements the lattice values were used \cite{RBC:2020kdj} and the BSM matrix elements were taken from \cite{Aebischer:2018rrz}. The master formula was then later generalized in \cite{Aebischer:2021hws} to include also next-to-leading order (NLO) QCD running effects. These were based on the NLO analysis for general $\Delta F=1$ effects in \cite{Aebischer:2021raf}. An interesting observable that has not yet been studied within the SMEFT is the $\Delta I = 1/2$ rule, which is closely related to $\epsilon'/\epsilon$ and which might provide a more refined picture of possible NP scenarios.

\section{$\Delta S=2$ transitions}

Also for the $\Delta S = 2$ observables a master formula exists, which was derived for general $\Delta F =2$ transitions in \cite{Aebischer:2020dsw}. In particular the mixing amplitude $[M_{12}]_{\text{BSM}}$ of the interaction Hamiltonian was determined, including renormalization group running above and below the EW scale. Due to the high sensitivity to NP effects even dimension-eight effects can be tested using $M_{12}$. Such an analysis was for instance performed in \cite{Liao:2024xel}, where the leading effects of operators with mass dimension eight were taken into account. A recent analysis that uses the constraining power of $\Delta M_K$ to test valid diquarks in the SMEFT context can be found in \cite{Englert:2024nlj}. 

Furthermore, for the $\Delta F =2$ case, several NLO effects are known within the SMEFT. For instance, the two-loop QCD and $y_t$ running in the SMEFT was studied in \cite{Aebischer:2022anv}. The one-loop matching from the SMEFT onto the Weak Effective Theory (WET) at EW scale was performed in \cite{Aebischer:2015fzz,Endo:2018gdn}. 

Another important effect when studying NP effects in the SMEFT, in particular for $\Delta S = 2$ transitions, is the concept of back-rotation. It was first discussed in \cite{Aebischer:2020mkv} in the context of Kaon physics and was later worked out in detail in \cite{Aebischer:2020lsx}. It describes the fact that for a given assumption of the flavor basis at a particular scale, after RG running the Yukawa matrices and other flavor matrices have to be rotated back to the initial basis via flavor rotations, since flavor assumptions are not stable under RG evolution.  

\section{Summary}

Kaon physics in combination with the SMEFT approach consists an important tool in the search for NP effects. In this regard rare Kaon decays play an important role when scrutinizing the NP parameter space, but also $\epsilon'/\epsilon$ and $\Delta S=2$ observables are valuable players in this game. Several observables are already worked out up to a high precision within the SMEFT, and various master formulae exist that include SMEFT and WET running as well as effects from hadronic matrix elements. Other observables like for instance the $\Delta I = 1/2$ rule are still unexplored and might be interesting to study in the future.

\acknowledgments 
J.A. is grateful for the invitation and for the funding from the Swiss National Science Foundation (SNSF) through grant TMSGI2-225951.

\bigskip


\end{document}